\documentclass[twocolumn,showpacs,prb,aps,epsf,amsmath,amssymb]{revtex4}

\usepackage{graphicx}% Include figure files
\usepackage{amsmath}

\begin{document}

\title{Single-valley high-mobility (110) AlAs quantum wells with anisotropic mass}

\author{S.~Dasgupta\footnote{dasgupta@wsi.tum.de}, S.~Birner, C.~Knaak, M.~Bichler, A.~Fontcuberta i Morral, G.~Abstreiter}
\affiliation{
\centerline{Walter Schottky Institut, Technische Universit\"at M\"unchen, Garching, D-85748 Germany}  
}
\author{M.~Grayson\footnote{Corresponding author: m-grayson@northwestern.edu}} 
\affiliation{
\centerline{Department of Electrical Engineering and Computer Science, Northwestern University, Evanston, IL 60208 USA}
\centerline{and}
\centerline{Walter Schottky Institut, Technische Universit\"at M\"unchen, Garching, D-85748 Germany}
}

\begin{abstract}

We studied a doping series of (110)-oriented AlAs quantum wells (QWs) and observed transport evidence of single anisotropic-mass valley occupancy for the electrons in a 150 \AA~wide QW. Our calculations of strain and quantum confinement for these samples predict single anisotropic-mass valley occupancy for well widths $W$ greater than 53 \AA. Below this, double-valley occupation is predicted  such that the longitudinal mass axes are collinear. We observed mobility anisotropy in the electronic transport along the crystallographic directions in the ratio of 2.8, attributed to the mass anisotropy as well as anisotropic scattering of the electrons in the X-valley of AlAs. 

\end{abstract}


\maketitle

Strain and quantum confinement play an important role in tuning the valley degeneracy in indirect bandgap semiconductors.\cite{sun,dhar} Two dimensional electron systems (2DESs) \cite{shayegan,dasgupta} as well as one dimensional systems (1D) \cite{moser1,moser2,gunawan} in AlAs have been previously studied and analyzed for (001)-oriented quantum wells (QWs), both of which have a reduced valley degeneracy from the bulk. The 2D systems are seen to have an isotropic mobility resulting from doubly degenerate valleys with orthogonally oriented anisotropic masses. But the other facets of growth, like the (110)-oriented QWs have not been extensively explored. The knowledge of the doping efficiency for this orientation is necessary to grow optimally doped cleaved-edge overgrown quantum wires \cite{moser1,moser2} and the valley degeneracy is expected to be different from (001)-oriented wells. In this paper, we present an experimental study of a doping series of double-sided-doped QWs grown in (110)-orientation and show experimental evidence of single-valley occupancy and an anisotropic electron mass for these QWs. This has been further complemented with effective mass calculations with finite barrier that take strain into account, which explain how for (110)-oriented AlAs QWs, single-valley occupation is expected for the square well width $W$ $>$ 53 \AA. Also, we deduced the donor binding energy and the doping efficiency.  Previously, anisotropic single-valley systems have only been investigated in piezo-strained (001) AlAs samples \cite{vaki,padm}. In contrast, the (110) AlAs samples investigated in this Letter are singly-degenerate as a result of the growth orientation, require no additional piezo sample preparation to reach single-valley occupancy, and tend to show higher mobilities than those reported in piezo-strained samples.

\begin{figure}[!ht]
\centering
\includegraphics[width=\linewidth,keepaspectratio]{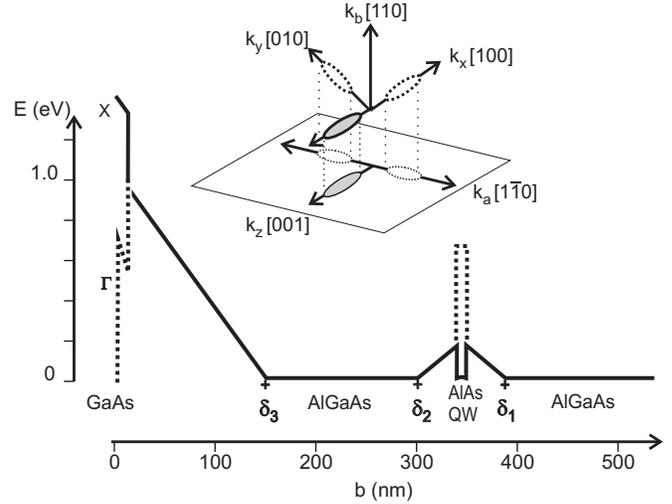}
\caption{(a) Energy band structure of the triple delta-doped (110) AlAs QW. The vertical axis denotes the energy scale, the + denotes the Si-delta doping layers in the AlGaAs layers, the solid line shows the X-point band, and the dashed line shows the $\Gamma$-point band. The structure is same as described in Ref.~4. (Inset) Electron occupation in a (110)-oriented QW, in momentum space. The $b$-axis denotes the growth axis, and there is a single in-plane valley with anisotropic mass occupied shown as shaded ellipsoid. Strain is measured in the $k_a, k_b, k_z$ growth-axis basis, and valley occupancy in the $k_x, k_y, k_z$ crystal-axis basis.}
\label{fig:fig1}
\end{figure}

Bulk AlAs is an indirect band gap III-V semiconductor with three degenerate conduction band valleys at the X-points of the Brillouin zone edge. The mass of each valley is anisotropic, with heavy longitudinal and light transverse mass $m_{\mathrm l}$ = 1.1 $m_{\mathrm e}$ and $m_{\mathrm t}$ = 0.2 $m_{\mathrm e}$, respectively \cite {shayegan,gun2,gun3}. In a 2D-confined system, the number of degenerate X-points depends on the growth direction and the QW width $W$ due to a balance of strain and confinement effects. For (001)-oriented QWs studied elsewhere with $W <$ 55 \AA, a single X-valley is occupied whereas $W >$ 55 \AA~yields doubly-degenerate occupied X-valleys in the QW.\cite{vandest}

We determine in this work how (110) QW valley degeneracy should depend on well width. The energy $E_{\tau}(\bf{k})$ of an electron in a valley $X_\tau$ with index $\tau$ (= $x, y$ or $z$) is
\begin{equation} 
\label{equ:EnergySplitting}
 E_{\tau}({\bf{k}}) = E_{\mathrm{kin}}({\bf k}) + E_{0,\tau} + \Delta_{\tau}({\bf k})
\end{equation}
where $E_{\mathrm{kin}}$ is the in-plane kinetic energy, $\bf k$ is the 2D in-plane momentum relative to the $\tau$-valley minimum, $E_{0,\tau}$ is the ground confinement energy, and $\Delta_\tau$ is the strain induced energy shift at the Brillouin zone edge at the X-point.

It is useful to introduce the cubic crystal axes $\vec{\bf x} = (x,y,z)$ and the growth axes $\vec{\bf a} = (a,b,z) = {\bf R} \vec{\bf x}$,\cite{moser1} related by the rotational transformation (Fig. 1) 
\begin{equation} 
\label{R}
 {\bf R} = \left[ \begin{array}{ccc} \frac{1}{\sqrt{2}}&\frac{-1}{\sqrt{2}}&0 \\ \frac{1}{\sqrt{2}} & \frac{1}{\sqrt{2}} &0 \\ 0 & 0 & 1 \end{array} \right].
\end{equation}
Confinement energies are best determined by transforming the mass tensor to the growth-basis $\vec{\bf a}$. In the crystal-basis $\vec{\bf x}$, the mass tensor $({\bf m}^{\tau}_{ij})^{-1}$ for the $\tau^\mathrm{th}$ valley is diagonal with ${\bf m}^{\tau}_{\tau \tau} = m_{\mathrm{l}}$ and ${\bf m}^{\tau}_{ii} = m_{\mathrm{t}}$ for $i \neq \tau$.  Transforming to the growth-basis $\vec{\bf a}$, the mass tensor becomes $({\bar{\bf m}}^\tau)^{-1} = {\bf R (m^\tau)^{-1} R^{-1}}$, so for the X$_x$-valley, for example with $\tau = x$
\begin{equation} 
\label{equ:MassTensorX}
 {(\bar{\bf m}^x)}^{-1} = \left[ \begin{array}{ccc} \frac{1}{m_d}&\frac{1}{m_f}&0 \\ \frac{1}{m_f} & \frac{1}{m_d} & 0 \\ 0 & 0 & \frac{1}{m_t} \end{array} \right]
\end{equation}
with diagonal $m_d = 2\frac{m_{\mathrm l} m_{\mathrm t}}{m_{\mathrm l} + m_{\mathrm t}}$ and off-diagonal $m_f = 2 \frac{m_{\mathrm l} m_{\mathrm t}}{m_{\mathrm t} - m_{\mathrm l}}$ mass terms.  The diagonal mass will enter into the Schr\"odinger equation for the quantum well confinement energy.

Deformation potential energies are best determined by transforming the strain tensor to the crystal-basis $\vec{\bf x}$. We assume that the AlAs QW ($a_{\mathrm {AlAs}} = 5.65252$ \AA)~is strained relative to the GaAs substrate lattice ($a_{\mathrm {GaAs}} = 5.64177$ \AA)~in agreement with previous publications. \cite{car,vande,vanKest}For (110) biaxial strain, the perpendicular component can be deduced from the in-plane strain as $\epsilon_{\perp} = -D\cdot\epsilon_{||}$, where $D = 0.6165$ is a constant that depends on the interface orientation and on the elastic constants of AlAs.\cite{vande,kri} In the growth-basis $\vec{\bf a}$, $(0,b,0)$ is the growth direction, and the strain tensor $\bar{\epsilon}_{ij}$ is diagonal with $\bar{\epsilon}_{aa} = \bar{\epsilon}_{zz} = \epsilon_{||}$, and $\bar{\epsilon}_{bb} = \epsilon_{\perp}$. Transforming to the $\vec{\bf x}$-basis, the strain tensor becomes $\epsilon = {\bf R}^{-1} \bar{\epsilon} {\bf R}$ with diagonal terms $\epsilon_{xx} = \epsilon_{yy} = \frac{\epsilon_{\perp} + \epsilon_{||}}{2}$ and $\epsilon_{zz} = \epsilon_{||}$.\cite{vande,kri} The energy at the $\tau^\mathrm{th}$ valley minimum shifts from the unstrained case by an amount $\Delta^\tau = \Xi \epsilon_{\tau \tau}$, where $\epsilon_{\tau \tau}$ is defined in the crystal-basis $\vec{\bf x}$, and the deformation potential for AlAs is $\Xi = 6.11$ eV. \cite{vande} In our case, the single and double-degenerate X band edges are separated by 9 meV. Combining these new band edges with knowledge of the mass tensor, the Schr\"odinger equation can now be solved.  All this information is automatically included in the publicly available simulation software nextnano$^3$ (\cite{nextnano}) yielding a cross-over width $W_{\mathrm c}$ = 53\AA.

For $W > W_\mathrm c$, there is only a single occupied in-plane valley shown in shaded grey in Fig.~1 inset. The two higher-energy out-of-plane valleys are shown as dotted ellipsoids. For $W < W_\mathrm c$ the role reverses, and these out-of-plane valleys become occupied. The projection of these valleys would be the white ellipsoids in the plane of the QW. Hence, the two valleys would have collinear longitudinal mass axes, in contrast to the (001) QW case where they are orthogonal.

Samples were grown on (110) GaAs substrates using molecular beam epitaxy. The structure of the samples is similar to the one discussed in Ref.~4 for the (001) facet and has been shown in Fig.~1. There is a 150 {\AA} wide QW with three Si $\delta$-doping layers. $\delta_1$ and $\delta_2$ separated from the QW by Al${_{0.45}}$Ga${_{0.55}}$As spacers provide electrons to the QW. These two $\delta$-doping layers are doped equally with a Si density $n_{\delta_1}$ = $n_{\delta_2}$ = $n_{\mathrm {Si}}$. The Si doping near the surface has higher density $n_{\delta_3}$ = $2.7n_{\mathrm {Si}}$ and satisfies the surface states to pin the conduction band to the donor binding energy upon saturation. Various samples were grown with different doping $n_{\mathrm {Si}}$ indexed G through M.\cite{sample}

\begin{figure}[!ht]
	\includegraphics[width=\linewidth,keepaspectratio]{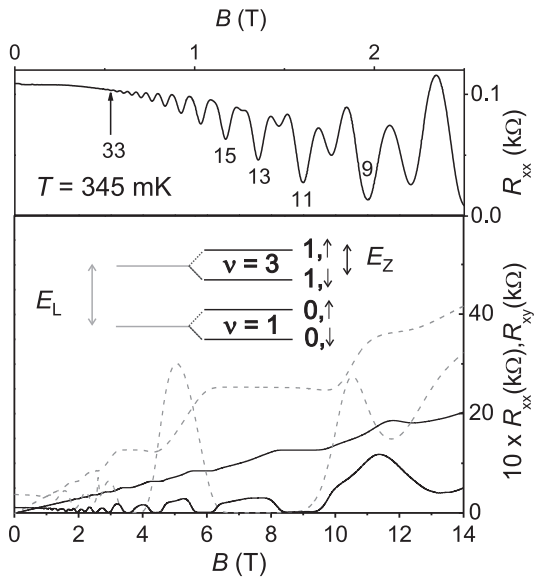}
	\caption{Typical $R_\mathrm {xx}$ and $R_\mathrm {xy}$ traces at 345 mK in the dark (dashed) and post-illumination (solid) for sample J in the van der Pauw configuration. In the dark, $n = 1.8\times 10^{11}$ cm$^{-2}$ and $\mu$ = $4.0\times 10^{4}$ cm$^{2}$/Vs and post-illumination, $n = 3.5\times 10^{11}$ cm$^{-2}$ and $\mu$ = $5.3\times 10^{4}$ cm$^{2}$/Vs. Inset on top shows the post-illumination longitudinal resistance for sample L at low magnetic fields. The odd periodicity of the prominent minima, is one indication of single-valley occupancy. The ladder diagram in the inset shows the Landau level splitting $E_L$ and Zeeman splitting $E_Z$ for $g^*m^* = 1$.  The prominence of odd integer gaps suggests $g^*m^* >1$ in the present system.}
	\label{fig:graph2}
\end{figure}

Indium contacts were annealed at 450 $^{\circ}$C for 100 s. Two-point contact resistance was around 100 k$\Omega$ at 300 K and around 40 k$\Omega$ at 4.2 K. Samples were illuminated using a red LED of wavelength 635 nm. We performed post-illumination anneal at around 25 K (PIA) for these samples, described previously for (001) AlAs samples,\cite{dasgupta} to obtain persistent photoconductivity. Typical longitudinal ($R_{xx}$) and transverse ($R_{xy}$) resistance variation with magnetic field at 330 mK in the dark and post-illumination for sample J in van der Pauw geometry is plotted in Fig.~2. The density of the samples in subsequent figures were deduced from such measurements. 

The first evidence of single-valley occupancy for (110)-orientation comes from the $\nu = 2n + 1$ periodicity at low magnetic field Shubnikov de Haas oscillations for the PIA data (Fig.~2 top).  %, compared to the 4-fold periodicity seen in double-valley degenerate systems. 
Due to the heavy cyclotron mass in AlAs, $m^* = (m_{\mathrm t}m_{\mathrm l})^{1/2} = 0.47 m_{\mathrm e}$ and large Lande g-factor of $g^* =$ 2 (Ref.~21), the bare Zeeman energy $E_{\mathrm Z}$ is about half the bare cyclotron energy $E_{\mathrm L}$, so that $m^*g^* \sim 1$ in the absence of interactions, the case shown in the Fig.~2 inset. Exchange and correlation enhancement of $m^*g^* \sim 1$ is known to occur in AlAs quantum Hall systems\cite{padm}, and for the range $1 < m^*g^* < 2$, the Zeeman gap will be larger than the cyclotron gap and odd filling factor $\nu = 2n + 1$ will dominate at low fields.  As shown in the data of Fig.~2 top, the odd integers clearly dominate $\nu = 9$ upwards and persist to filling factors as high as $\nu = 33$.  In double-valley systems, by comparison, an additional factor of 2 appears in the filling factor due to valley degeneracy, and one observes prominent gaps in the series $\nu = 4n + 2$\cite{dasgupta}.

The second evidence of single-valley occupancy comes from the anisotropic mobility in (110) QWs.  We performed mobility measurements on a L-shaped Hall bar fabricated along the crystallographic axes [001] and [1$\bar{1}$0]. The density in both arms is found to be the same, n = $1.65\times 10^{11}$ cm$^{-2}$ and the mobilities in the two directions are $\mu_{[001]}$ = $0.3\times 10^{5}$ cm$^{2}$/Vs and $\mu_{[1\bar{1}0]}$ = $1.0\times 10^{5}$ cm$^{2}$/Vs. The mobility anisotropy is consistent with anisotropic mass electrons conducting in a single-valley with the heavy mass oriented along the [001] direction and the light mass along the [1$\bar{1}$0] direction, as expected in the single-valley degenerate system. We note that a naive Drude model of mobility is not sufficient since the mobility ratio is $\frac {\mu_{[1\bar{1}0]}} {\mu_{[001]}} = 2.8$, whereas the inverse mass ratio is $\frac {m_{[001]}} {m_{[1\bar{1}0]}} = \frac {m_{\mathrm l}} {m_{\mathrm t}}= 5.5$. There have been previous publications reporting similar non-Drude effects in anisotropic mass systems\cite{bishop}, and like these authors we propose that anisotropic scattering is responsible.\cite{tok} 

When the mobility is measured using van der Pauw geometry one obtains the mobility, $\mu_{\mathrm {VdP}}$,\cite{vdp1} which is measured to be $0.40\times 10^{5}$ cm$^{2}$/Vs. This can be related to the mobility in each crystallographic direction as ${{\mu^{'}}_{\mathrm {VdP}}} = {(\mu_{xx}\mu_{yy})}^{1/2}$.(Ref.~22) For our sample, the mobility was found to be ${{\mu^{'}}_{\mathrm {VdP}}} = 0.50\times 10^{5}$ cm$^{2}$/Vs. This is within reasonable agreement to the van der Pauw mobility, which is within 20$\%$ of the average value. Comparing the (110) results here with Ref.~2 which has an identical structure on the (001) facet, we note approximately a factor of 6 reduction in van der Pauw mobility for the (110) grown quantum wells. These results would be discussed in more detail in an upcoming publication.\cite{dasgupta2}

\begin{figure}[!ht]
\includegraphics[width=\linewidth,keepaspectratio]{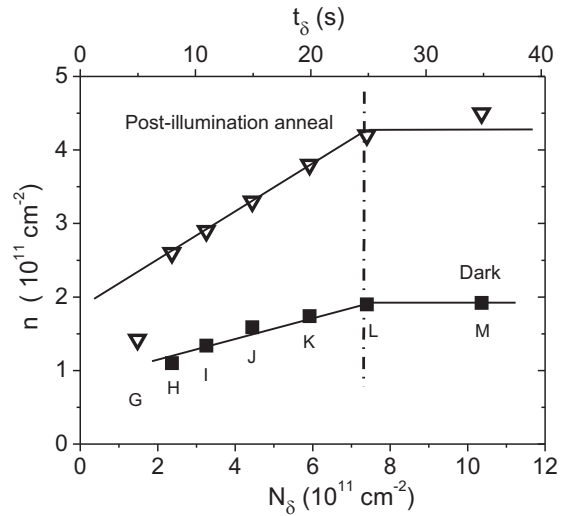}
\caption{The density of two dimensional electron system in a (110)-oriented AlAs quantum well as a function of the Si delta doping $n_{\mathrm {Si}}$. The top $x$-axis defines the time of Si doping corresponding to the doping densities. The vertical dashed line shows the saturation threshold $n_{\mathrm{sat}}$ for both the dark and the post illumination(red LED $\triangledown$) conditions. The solid lines are plotted as an aid to locating the saturation threshold.}
	\label{fig:graph3}
\end{figure}

In Fig.~3 the 2DES density of the samples are plotted as a function of different Si doping concentrations $n_{\mathrm {Si}}$ deduced from the doping times, $t_{\mathrm {Si}}$, with a calibrated Si flux of $2.9\times 10^{10}$ cm$^{-2}$ s$^{-1}$ for 11.4 A heater current. The dark electron density is seen to saturate at around $n_{\mathrm {DK}}$ = $1.9\times 10^{11}$ cm$^{-2}$. Post-illumination anneal (PIA) density, plotted as triangles, is seen to saturate at about $n_{\mathrm {PIA}}$ = $4.2\times 10^{11}$ cm$^{-2}$. Sample L represents the optimally doped sample, since at higher doping approximiately the same saturation density $n_{\mathrm {PIA}}$ is observed. Sample G works only post illumination. In the lowest density sample G, the post-illumination density drops abruptly, most likely indicating that the surface $\delta$-doping layer is no longer in saturation. From the experimental dark density of $n_{\mathrm {DK}}$, we deduce a donor binding energy of $\Delta_{\mathrm {DK}} =$ 66 meV and for $n_{\mathrm {PIA}}$, we deduce the post-illumination saturation binding energy of $\Delta_{\mathrm {PIA}} =$ 0 meV using the same analysis presented in a previous publication.\cite{dasgupta} The doping efficiency for the samples in the dark and post-illumination was calculated using the doping density $n_{\mathrm {Si}}$ at the saturation threshold shown with a vertical dashed line. We define from Ref.~4,
\begin{equation}
\label{equ:doping efficiency}
	\eta_{\mathrm {DK, PIA}} = \frac{n_{\mathrm {DK, PIA}}}{2 n_{\mathrm {Si}}}
\end{equation}
where $\eta$ is the doping efficiency. The factor of 2 in the denominator arises from the double-sided $\delta$-doping layers, assuming that all surface states have been screened by the top $\delta$-layer. Using this equation, we obtain the doping efficiency of $\eta_{\mathrm {DK}}= 7\%$ in the dark and $\eta_{\mathrm {PIA}} = 15\%$ post-illumination anneal.  

In summary, we have shown experimentally the occupation of the single X-valley by the 2DES in AlAs QWs in this (110)-orientation, which shows anisotropic mobility that can be partly attributed to the mass anisotropy. We have also presented results of a model calculation for the cross-over width in AlAs for (110)-orientation, which defines the QW width below which double-valleys are occupied and above which a single-valley is occupied. Furthermore, we have determined the binding energy of Si in $\delta$-doped layers in Al${_{0.45}}$Ga${_{0.55}}$As in the dark and after illumination for the (110)-orientation and found them to be in the range of the values found for (001)-orientation. The doping efficiency for the Si $\delta$-layers has been calculated to be $7\%$ in the dark and $15\%$ post-illumination anneal. These parameters will be instrumental in optimizing mobility in (110)-oriented AlAs QWs and cleaved-edge overgrowth structures.
\\
\\
This work was funded by BMBF nanoQUIT Project 01BM470 and the Nanosystems Initiative Munich (NIM). 

\end{document}